\documentclass[letterpaper,english,aps,prl,floatfix,twocolumn,showpacs,amsfonts,amssymb]{revtex4}

\usepackage{graphicx}
\usepackage{amssymb}
\usepackage{overpic}

\usepackage{babel}

\newcommand{\beq}{\begin{equation}}
\newcommand{\eeq}{\end{equation}}
\newcommand{\bea}{\begin{eqnarray}}
\newcommand{\eea}{\end{eqnarray}}

\begin{document}

\title{Functional Renormalization Group Analysis of the
  Half-filled One-dimensional Extended Hubbard Model}

\author{Ka-Ming Tam, Shan-Wen Tsai\footnote{Present address: Department of
    Physics, University of California, Riverside, CA 92521, USA.}, and David K. Campbell}

\affiliation {Department of Physics, Boston University, Boston, MA 02215}

\date{\today}

\begin{abstract}

We study the phase diagram of the half-filled one-dimensional extended
Hubbard model at weak coupling using a novel functional renormalization group
(FRG) approach. The FRG method includes in
a systematic manner the effects of the scattering processes involving
electrons away from the Fermi points. Our results confirm the existence of
a finite region of bond charge
density wave (BCDW), also known as a ``bond order wave'' (BOW),
 near $U = 2V$ and
clarify why earlier 
g-ology calculations have not found this phase. We argue that this is an
example in
which formally irrelevant corrections change the topology of the phase diagram.
Whenever marginal terms lead to
an accidental symmetry, this generalized
FRG method may be crucial to characterize the phase diagram accurately.



\end{abstract}
\maketitle



The one-dimensional extended Hubbard model (EHM) has been
studied extensively for many years,
both because of its rich phase diagram \cite{Baeriswyl} and because
of its possible applications to quasi-1D organic crystals \cite{organics} and
conducting polymers \cite{polymers}. Despite this long history, a
controversy has recently arisen concerning the possible existence of a bond 
order charge
density wave (BCDW, also called a ``bond order wave'' (BOW))
phase separating the well-known spin density wave (SDW) and
charge density wave (CDW)
phases of the EHM at half-filling. This phase has been suggested by
Nakamura \cite{Nakamura} and supported by quantum Monte Carlo
(QMC) \cite{Sengupta} and more recently density matrix renormalization group
(DMRG)\cite{Zhang}
calculations. However, this phase was not obtained in earlier numerical and
analytical work \cite{Emery,Solyom,Cannon,Dongen,Hirsch1,Voit,Jeckelmann}. In
particular, this phase
is absent in standard one-loop g-ology \cite{Emery,Solyom} and
bosonization \cite{Voit} calculations. This disagreement poses a serious
question, which we elucidate here. We reconcile the recent numerical
results \cite{Sengupta,Zhang} with g-ology by introducing a functional
generalization of the standard g-ology formalism. Our functional
renormalization group (FRG) method offers a consistent and
well-controlled approximation that predicts a finite region in parameter
space in which the BCDW
phase is spontaneously formed for the half-filled EHM at weak coupling.
Our results thus go beyond the important earlier study of Tsuchiizu and
Furusaki \cite{TF}, who were able to obtain the BCDW
phase with a standard RG using {\it ad hoc} approximations.

The Hamiltonian of the EHM is given by
\begin{eqnarray}
\label{Hamiltonian}
   H &=& - t \sum_{i,\sigma}(c_{i+1,\sigma}^{\dagger}c_{i,\sigma}
    + H.c.) + U\sum_{i}n_{i,\uparrow}n_{i,\downarrow}\nonumber\\
      & & + V\sum_{i}n_{i}n_{i+1} - \mu \sum_{i} n_i,
\end{eqnarray}
where $t$, $U$, and $V$ are the nearest-neighbor hopping, on-site
interaction, and the nearest-neighbor interaction, respectively.
Here we study the EHM at half-filling ($\mu = 0$). It is well
established that for repulsive interactions ($U, V > 0$), the
system is in a CDW phase for large values of $V/U$ and in a SDW
phase for small $V/U$. Weak-coupling RG studies \cite{Emery,Solyom}
find the boundary between these two phases to be at $U=2V$. Early
strong-coupling numerical studies \cite{Hirsch1} and higher-order
perturbation theory \cite{Dongen} have found the phase boundary to
be slightly shifted away from the $U=2V$ line, with a larger SDW
phase. Stochastic series expansion QMC studies \cite{Sengupta}
found that the BCDW phase exists in a finite region around the line
$U=2V$ and that it ceases to exist when the interaction exceeds
a critical value.
There are disagreements between the published DMRG results.
An earlier result \cite{Jeckelmann} showed that the BCDW exists
only precisely at the CDW/SDW phase boundary at intermediate couplings. A
more recent DMRG calculation \cite{Zhang} obtained the same phase
diagram as the QMC study \cite{Sengupta}.

The standard RG--''g-ology''-- has proven to be a powerful
method for studying  low-energy properties of
interacting one-dimensional systems at weak-coupling. By integrating out 
high-energy modes, one obtains flow equations for the marginal
couplings such as the two body interaction vertices.
These interaction vertices are in principle
functions of three momenta, which can take any value within the
Brillouin zone and correspond to the momenta of the incoming and
outgoing electrons. The fourth momentum is determined by momentum
conservation. In standard g-ology, the interaction processes are
classified according to the branch label (right or left) of the
electrons involved. All further dependence on the momenta, {\it i.
e.}, the dependence on the magnitude of the momenta, is neglected,
since only the dependence on the direction of the momenta is marginal.
The radial dependence is irrelevant according to scaling and power-counting
arguments \cite{Shankar}.  It is important to notice that irrelevant
operators renormalize to zero as the RG proceeds but may not be
small in the beginning of the flow. This is the key issue here, and
we will return to it later.

In g-ology, interaction processes are classified into backward
scattering ($g_1$), forward scattering involving electrons from
two branches ($g_2$) and from the same branch ($g_4$), and Umklapp
process ($g_3$).
As the scattering between electrons with the same spin can be obtained
from scattering between electrons with different spins
 \cite{Zanchi}, we shall ignore all spin indices, leaving it 
understood that all processes are between electrons with different
spins. For the EHM, the bare values of the couplings are $g_1 =
g_3 = U - 2V$ and $g_2 = g_4 = U + 2V$.

Exactly at $U = 2V$, both $g_1$ and $g_3$ are equal to zero,
and they remain zero under the RG flow,
resulting in a
massless theory for both the spin and charge sectors. This is the underlying
reason that conventional
weak-coupling calculations (both g-ology and bosonization) find a direct
transition between the SDW and the CDW phase exactly at $U = 2V$, where both
gaps vanish simultaneously.
An important insight was provided by Nakamura \cite{Nakamura} (and further
explored by Tsuchiizu and
Furusaki \cite{TF}), who observed that there is no symmetry
principle that enforces $g_1$ and $g_3$ to vanish simultaneously
and that higher-order corrections may lift this degeneracy and
thereby change the topology of the phase diagram \cite{Affleck}.
They then adopted an idea from Penc and Mila \cite{Penc} and
applied the following two-step procedure: for the high-energy part
of the band ($\Lambda > \Lambda_{cutoff}$), second-order
perturbation theory is performed to find corrections for the
couplings $g$; these values are then used as the initial
conditions for the RG procedure, which is performed for the low
energy part of the band ($\Lambda < \Lambda_{cutoff}$).
This is sufficient to generate a finite region of BCDW phase.
Clearly, this procedure is {\it ad hoc} and relies on an arbitrary
choice for $\Lambda_{cutoff}$ (which in \cite{TF} is chosen to be half the
total bandwidth). The subsequent RG results, in
particular the size of the BCDW region, depend on the choice of
$\Lambda_{cutoff}$ and hence do not definitively answer the
question whether the BCDW phase is intrinsic in EHM at
half-filling.

The virtue of the one-loop {\it functional} RG we develop and employ below
is that it captures the BCDW phase in a
systematic manner without {\it ad hoc} manipulations. The key
point is that, while we truncate the flow equations to order $g^2$
as in standard one-loop calculations, we maintain full momentum
dependence of the interaction vertices. So instead of solving the
RG flow equations for four couplings $g_1$, $g_2$, $g_3$ and
$g_4$, we write the {\it functional} RG equation for
$g(k_1,k_2,k_3)$, where $k_1$, $k_2$, $k_3$ can be anywhere in the
Brillouin zone. Although the radial dependence is formally irrelevant and the
correponding terms will eventually flow to zero, their effect may be finite
when the energy cutoff is near the band edge, thereby breaking the 
accidental degeneracy.
Other irrelevant terms, such as higher-order vertices, are absent
in the beginning of the flow and we neglect them altogether,
just as in standard g-ology.
We stress that our procedure should in general not
qualitatively change the phase diagram--irrelevant operators will remain
irrelevant--but may be crucial when an 
accidental degeneracy occurs. In this case very different phases
may appear.

\begin{figure}[bth]
\centerline{
\includegraphics*[height=0.12\textheight,width=0.32\textwidth,viewport=80
310 510 555,clip]{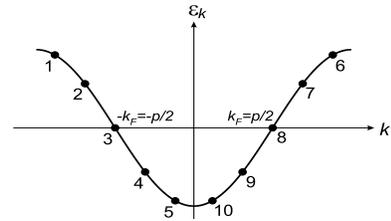}} \caption{Discretization of the momenta
in the Brillouin zone. This figure shows the case $N=10$.}
\label{fig:n}
\end{figure}

Our functional RG equations for the one-dimensional EHM at one-loop follow
closely the approach of Zanchi and Schulz \cite{Zanchi},
which itself is an adaptation of
the two-dimensional RG for fermions \cite{Shankar} to the case of
an arbitrary Fermi surface. This and other formulations of the
functional RG have recently been applied to several two-dimensional
interacting
electron systems \cite{Honerkamp1,Halboth,Tsai,Binz,Kampf}.
The crucial difference is that we consider
a finite number $N$ of divisions of the {\it magnitude} of the momenta,
while the two-dimensional calculations discretize the Fermi
surface into angular patches.  At
one-loop level our equations become:
\begin{eqnarray}
&&\frac{dg(k_1,k_2,k_3)}{d \Lambda} =
\nonumber\\
&-&\!\!\!\!\int\!d\underline{p} \frac{d}{d\Lambda}
[G_{\Lambda}(\underline{p})G_{\Lambda}(\underline{k})] g(k_1,k_2,k)
g(p,k,k_3)
\nonumber\\
&-&\!\!\!\!\int\!d\underline{p} \frac{d}{d\Lambda}
 [G_{\Lambda}(\underline{p})G_{\Lambda}(\underline{q_1})]
g(p,k_2,q_1) g(k_1,q_1,k_3)
\nonumber\\
&-&\!\!\!\!\int\!d\underline{p} \frac{d}{d\Lambda}
 [G_{\Lambda}(\underline{p})G_{\Lambda}(\underline{q_2})][-\!2g(k_1,p,q_2)g(q_2,k_2,k_3)
\nonumber\\
&+&\!\!g(p,k_1,q_2) g(q_2,k_2,k_3)\!+\!g(k_1,p,q_2)
g(k_2,q_2,\!k_3)],
\end{eqnarray}
where $k=k_1+k_2-p$, $q_1=p+k_3-k_1$, $q_2=p+k_3-k_2$,
$\underline{p} = (p,\omega)$, $\int d\underline{p}=\int
dp\sum_{\omega}1/(2\pi\beta)$, and $G_{\Lambda}$ is the propagator
with cutoff $\Lambda$.

The  high-energy modes are integrated from the full bandwidth
$\Lambda_0$ (both positive and negative) to $\Lambda$, towards the
Fermi surface. The cutoff $\Lambda$ is parameterized by the RG
parameter $\ell$ as $\Lambda = \Lambda_0 \exp(-\ell)$. The initial
condition for $g(k_1,k_2,k_3)$ is given by the Fourier transform
of the $U$ and $V$ interaction terms. A direct analytical
solution of the functional RG equation does not seem possible, and
we use numerical calculations for solving the coupled
integral-differential equations. For this purpose, the Brillouin
zone is divided into $N$ segments. Fig. \ref{fig:n} shows the
discretization scheme for $N=10$.

We can follow the flows of the couplings to determine the dominant
instability, but a more definitive answer is given by comparing the
susceptibilities corresponding to different broken symmetry
states. We consider the susceptibilities of SDW, CDW, BSDW, and
BCDW in the long-wavelength limit.
Their general form is
\begin{eqnarray}
\label{SDW_susceptibility}
\chi^{\delta}_{\Lambda}(\pi\!)\!\!=\!\!\!\int\!\!\! D(1,2)
f(p_{1}\!)f(p_{2}\!) \langle
c_{p_{1},\sigma_{1}}^{\dagger}c_{p_{1}\!+\!\pi,\sigma_{1}}
c_{p_{2}\!+\!\pi,\sigma_{2}}^{\dagger}c_{p_{2},\sigma_{2}}\!\rangle,
\end{eqnarray}
where $p_i$ is the momentum at energy $\xi_{i}$,
$\int D(1,2)\equiv \int_{|\xi_{1}|>\Lambda}d\xi_{1}J(\xi_{1})
\int_{|\xi_{2}|>\Lambda}d\xi_{2}J(\xi_{2})\sum_{\sigma_{1},\sigma_{2}}s_{\sigma_{1}}s_{\sigma_{2}}$,
and $J(\xi)$ is the Jacobian for the coordinate transformation
from $k$ to $\xi_k$. For $\delta = SDW$ and $\delta =
BSDW$:$s_\uparrow =1, s_\downarrow = -1$. For $\delta = CDW$ and
$\delta = BCDW$: $s_\uparrow =1, s_\downarrow = 1$. For $\delta =
SDW$ and $\delta = CDW$:$f(p)=1$. For $\delta = BSDW$ and $\delta
= BCDW$: $f(p)=\sin(p)$. 
In momentum space, the difference between
site
and bond ordering is just in the form factor, which is s-wave for site 
orderings and p-wave for bond orderings.
\begin{figure}[htb]
\centerline{
\includegraphics*[height=0.17\textheight,width=0.235\textwidth,viewport=62 230 500 550,clip]{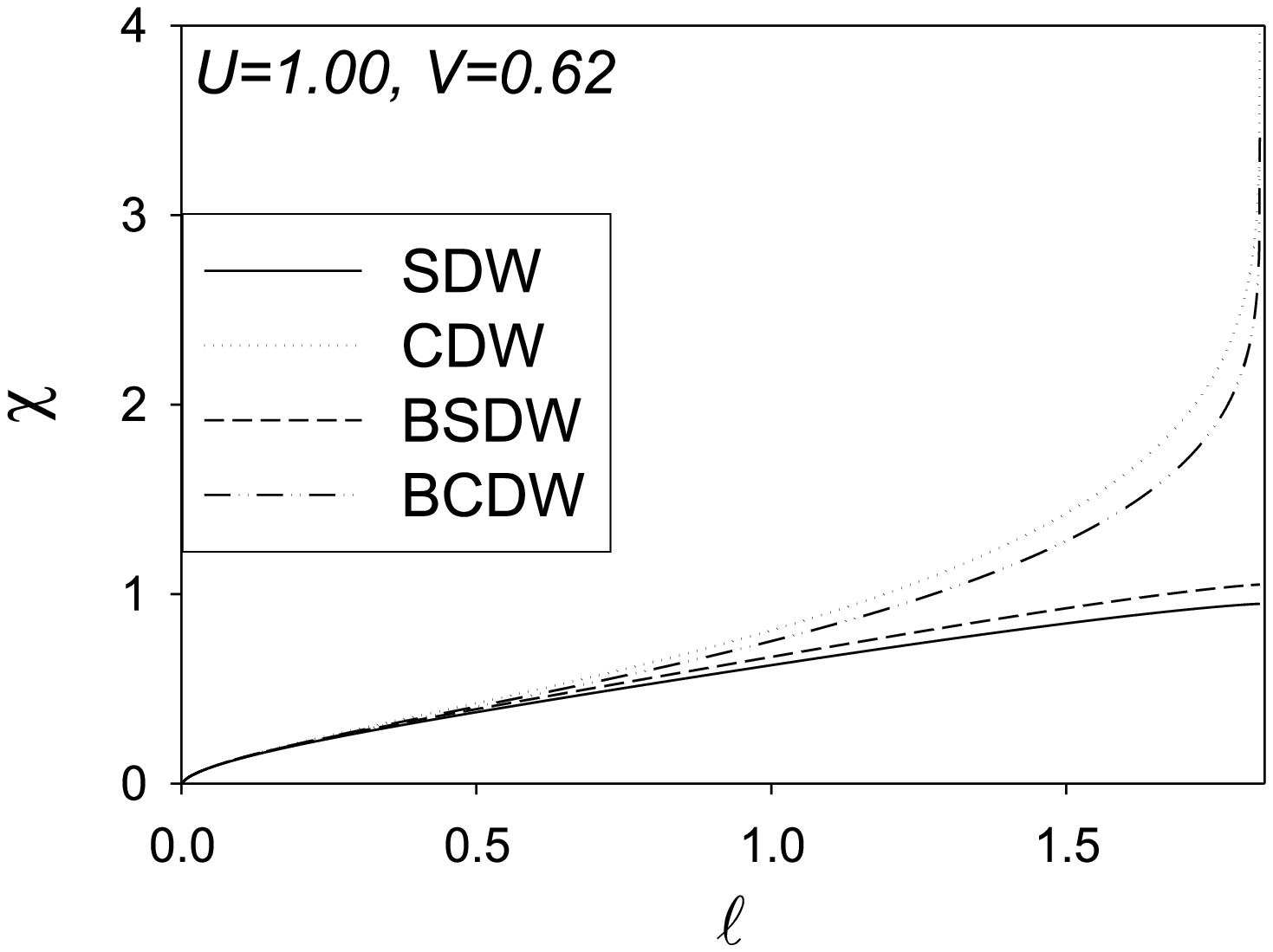}
\includegraphics*[height=0.17\textheight,width=0.235\textwidth,viewport=62 230 500 550,clip]{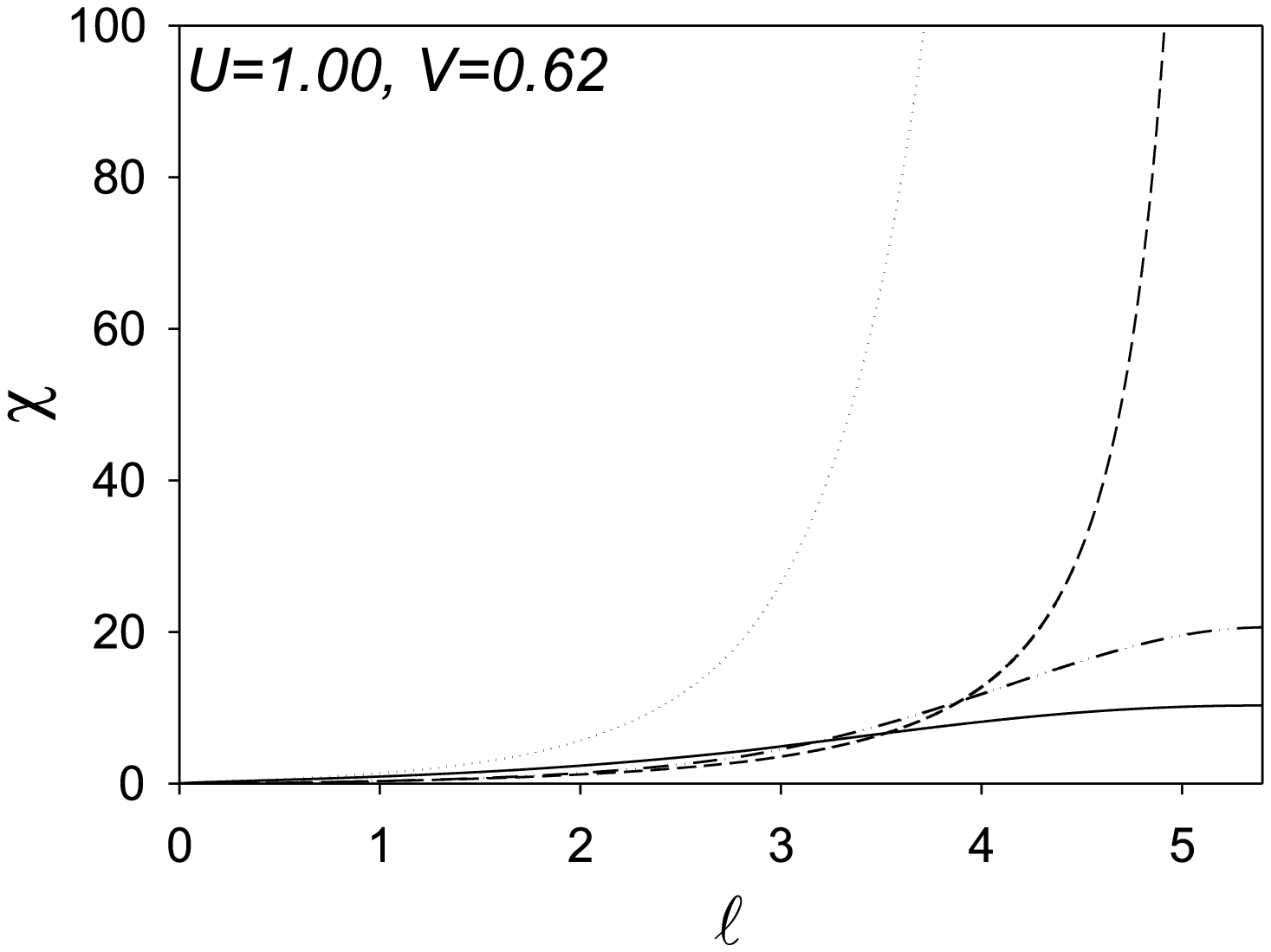}}
\centerline{
\includegraphics*[height=0.17\textheight,width=0.235\textwidth,viewport=62 230 500 550, clip]{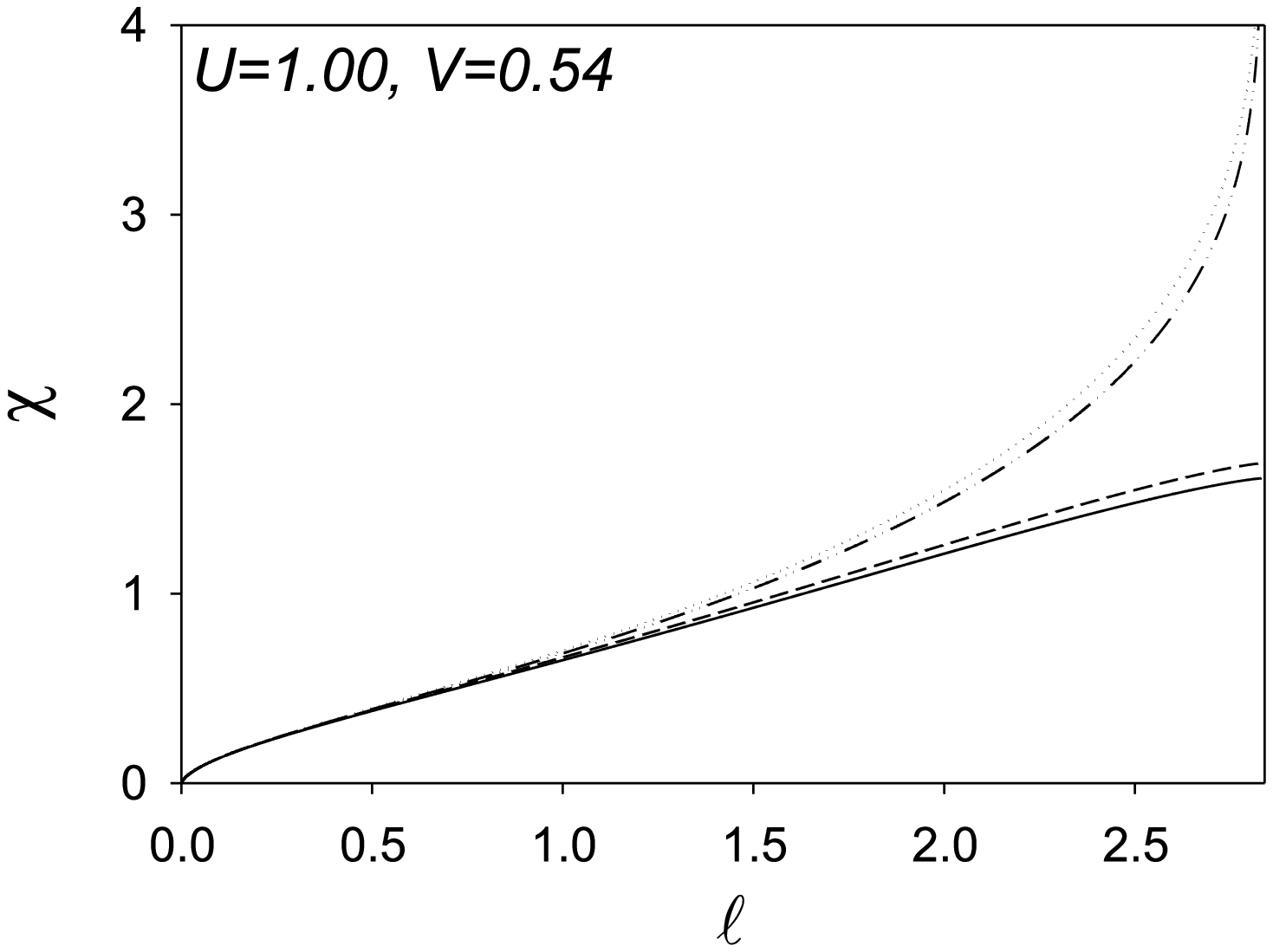}
\includegraphics*[height=0.17\textheight,width=0.235\textwidth,viewport=62 230 500 550,clip]{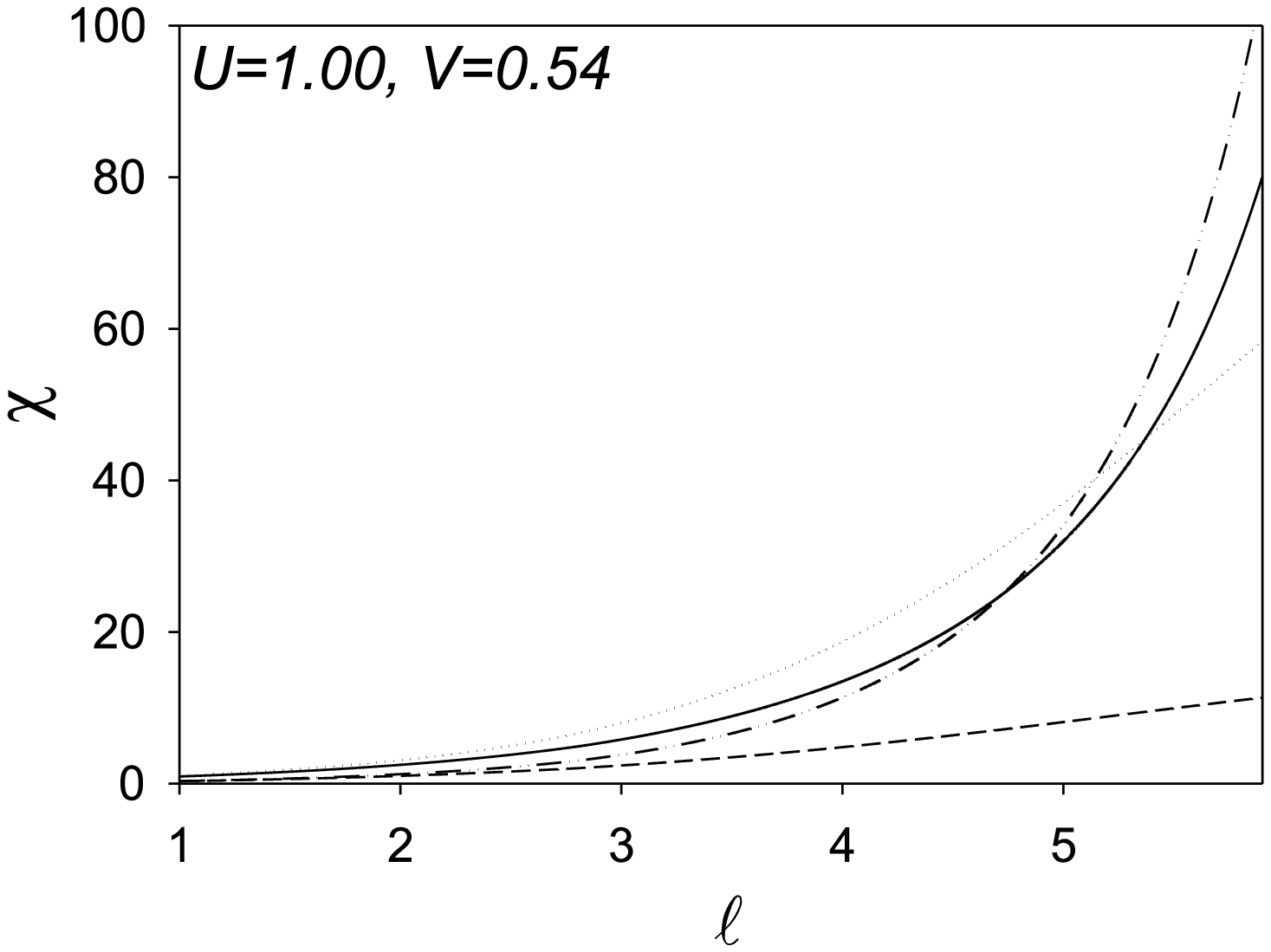}}
\centerline{
\includegraphics*[height=0.17\textheight,width=0.235\textwidth,viewport=62 230 500 550,clip]{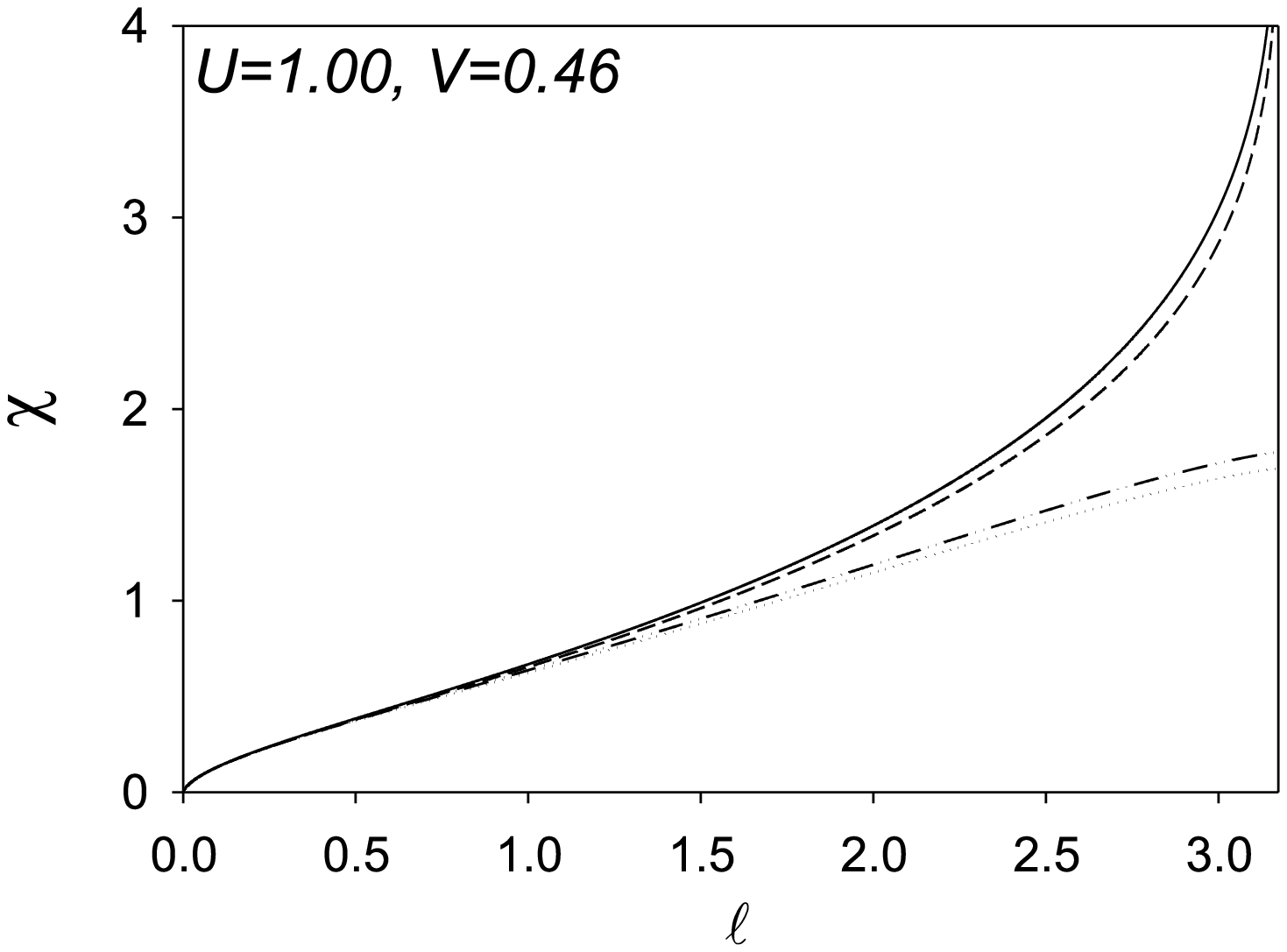}
\includegraphics*[height=0.17\textheight,width=0.235\textwidth,viewport=62 230 500 550,clip]{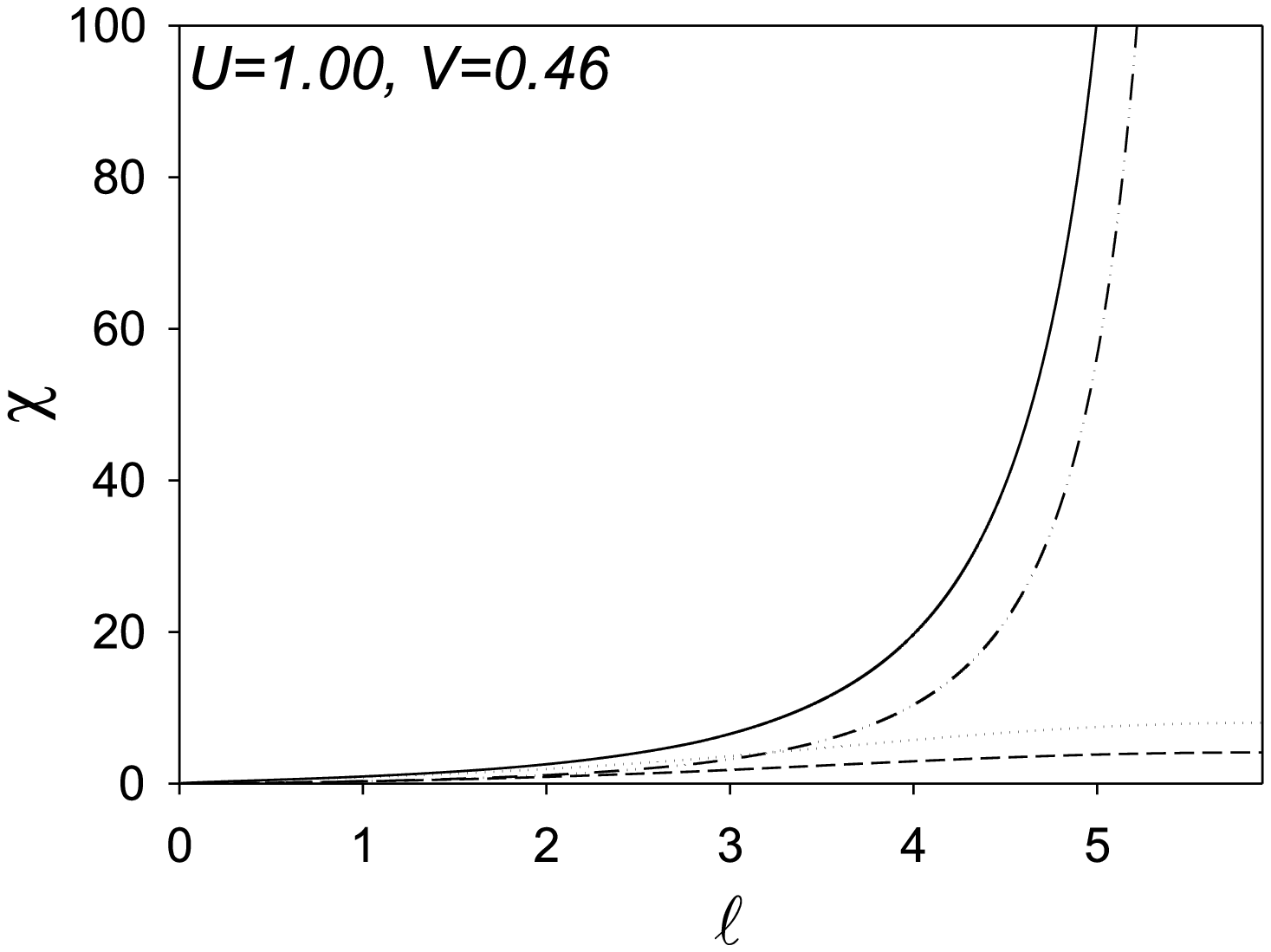}
} \caption{The flows of SDW (solid lines), CDW (dotted lines),
BSDW(dashed lines), BCDW(dotted-dashed lines) susceptibilities as
function of $\ell$ for $U=1$, and $V=0.46,0.54,0.62$. The left
column is for $N=2$, corresponding to standard g-ology, and the
right column is for $N=50$.} \label{fig:erptsqfit}
\end{figure}

Under the RG procedure, the susceptibilities also flow, with the
dependence on $\Lambda$
appearing both in the integration and in the flow of the expectation value
$\langle ... \rangle$.
The dominant instability is determined by the most
divergent susceptibility as $\ell$ is increased.
The RG equations for the
susceptibilities are,
\begin{eqnarray} \label{SDW_susceptibility RGE}
\frac{d \chi^{\delta}_{\Lambda}(\pi)}{d\Lambda}\!\!\!&=&\!
-\!\!\!\int\!\!\!d\underline{p} \frac{d}{d\Lambda}[
G_{\Lambda}(\underline{p})G_{\Lambda}(\tilde{\underline{p}})](Z^{\delta}_{\Lambda}(p))^2,  
\\
\frac{d Z^{\delta}_{\Lambda}(p)}{d\Lambda} \!\!\!&=&\!
\!\!\!\!\int\!\!\!d\underline{p}^{\prime}\frac{d}{d\Lambda}[
G_{\Lambda}(\underline{p}^{\prime})G_{\Lambda}(\tilde{\underline{p}^{\prime}})]
Z^{\delta}_{\Lambda}\!(p^{\prime})g^{\delta}\!(p^{\prime}\!,p).
\end{eqnarray}
where $\tilde{\underline{p}} = \underline{p} + (\pi,0)$. For $\delta=SDW$ and $\delta=BSDW$: $g^{\delta}(p^{\prime},p)=
-g(p+\pi,p^{\prime},p)$. For $\delta=CDW$ and $\delta=BCDW$:
$g^{\delta}(p^{\prime},p)=2g(p^{\prime},p+\pi,p)-g(p+\pi,p^{\prime},p)$.
The function $Z^{\delta}(p)$ is the effective vertex in the
definition for the susceptibility $\chi^{\delta}$. Its initial
condition is $1$ for SDW and CDW, and $\sin(p)$ for BSDW and BCDW.
The RG equations for susceptibilities are solved with initial
condition $\chi^{\delta}_{\Lambda=\Lambda_{0}}(\pi)=0$.

It is instructive to compare the difference in the
susceptibility flows for the (conventional) case of two
scattering points $N=2$ and
for cases of multiple scattering points $N\geq2$.
We focus on the case $U=1$ (in units of $t$) which is in the
weak-coupling regime ($U < \Lambda_0 = 2t$). In the left panels of
Fig. \ref{fig:erptsqfit} we show the flows of susceptibilities for
$N=2$ {\it i. e.} for standard g-ology. In the right panels are
results for functional RG with $N=50$. For each case, results are
shown for $V=0.46, 0.54$ and $0.62$. These three values were
chosen to cover the SDW, BCDW and CDW phases around the $U = 2 V$
line.

First, we note that for $N=2$, the susceptibilities tend to
diverge more quickly with $\ell$.
This is because, in effect,
all the renormalization corrections to the irrelevant couplings (involving
radial excursions away from the Fermi points) have been assigned to the
marginal ones ($g_1$, ..., $g_4$ of standard g-ology).
%
This significantly enhances the rate of increase of the couplings
and of the susceptibilities. Second, for $N=2$ the SDW dominates
for $V=0.46$ where $U\geq2V$, and CDW dominates when $U\leq2V$, as
can be seen for $V=0.54$ and $0.62$. For $N=2$ {\it all} the
density wave susceptibilities are degenerate at $U=2V$. Therefore,
for $N=2$ there is no finite region of BCDW phase.


Importantly, for $N>2$, we find that the BCDW suscetibility is dominant
in a finite range around $U=2V$. This is shown in
the Fig. \ref{fig:erptsqfit} for $N=50$, for the case $U=1,
V=0.54$. For smaller (larger) $V$ the system is in the SDW
(CDW) phase, as predicted by the standard g-ology. The pattern of
SDW-BCDW-CDW for increasing $V$ at a fixed $U$ can be obtained by
having only 6 scattering points along the band.

\begin{figure}[bth]
\centerline{
\includegraphics*[height=0.16\textheight,width=0.38\textwidth,
viewport=58 233 510 560,clip]{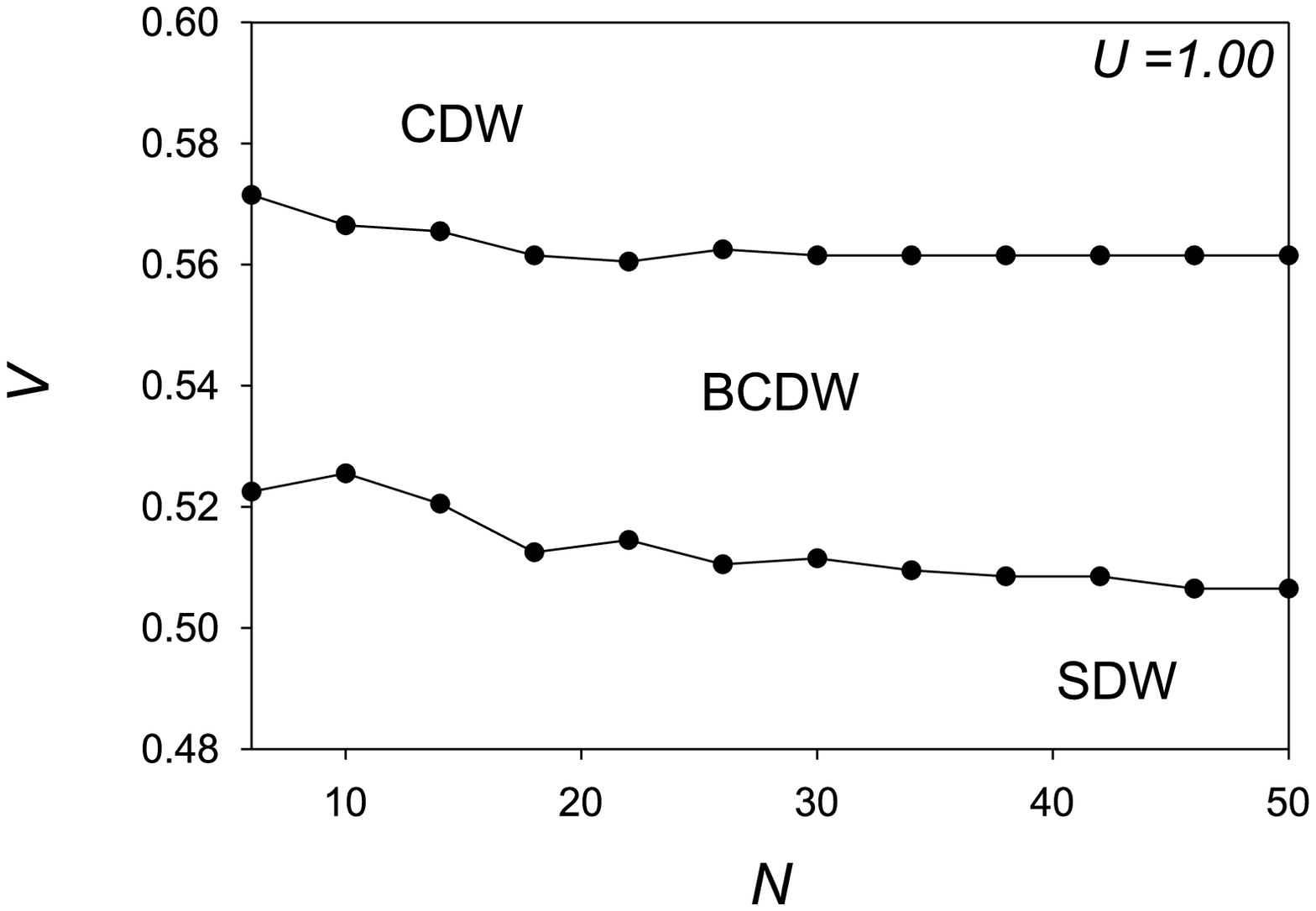}} \caption{The shifts
of the SDW-BCDW and BCDW-CDW phase boundaries with the number of
momentum divisions $N$. One can note that the phase boundaries
converge quickly with $N$, and they do not have any significant
change for $N > 30$.}
\label{fig:convergence}
\end{figure}

For our results to be reliable, it is important that they converge
to a fixed result as $N$ increases. Fig. \ref{fig:convergence} shows that the
phase boundaries converge quickly with $N$. Therefore, the $N=50$
results presented in Fig. \ref{fig:erptsqfit} have reached the
large $N$ limit.

\begin{figure}[bth]
\centerline{
\includegraphics*[height=0.16\textheight,width=0.44\textwidth,
viewport=20 230 600 560,clip]{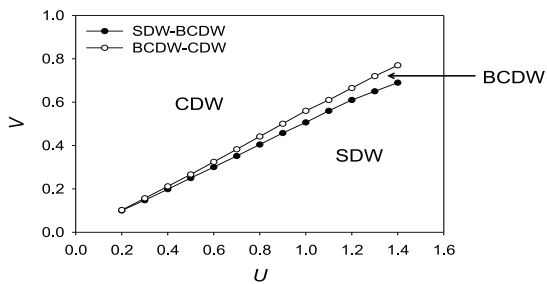}} \caption{The phase
diagram of one- dimensional EHM at half-filling in the weak
coupling regime.}
\label{fig:pd}
\end{figure}

The phase diagram we obtain is shown in Fig. \ref{fig:pd}. By
using the momentum-dependent functional RG, we have confirmed that the
BCDW phase extends to very weak coupling (at least down to $U
\approx \Lambda_{0}/10$) and expands with increasing $U$. This
regime is difficult to access via QMC studies, and there has also been a
controversy regarding the DMRG results in this limit.
Intriguing
phenomena in the strong-coupling regime, such as the shrinkage of the
BCDW phase 
can not be reliably 
studied by this weak-coupling FRG method. This method takes
into account irrelevant terms that couple spin and charge degrees of freedom,
which has been argued in Ref.\cite{Nakamura} to be the cause for the
disappearance of the BCDW phase at strong coupling. Nevertheless, the RG
expansion is only valid for weak couplings and we focus on this limit in the
present work. 

In summary, we have studied the phase diagram of the EHM at half-filling
using a generalization of the standard g-ology to a momentum-dependent,
functional RG. In conventional terms, our approach includes
formally irrelevant terms corresponding to interaction vertices
involving electrons at high momenta. In the present case,
our procedure changes the phase diagram {\it qualitatively}
because it breaks an accidental symmetry
between the backscattering ($g_1$) and the Umklapp ($g_3$)
processes at $U = 2V$. The full momentum dependence is included in a
systematic way by discretizing the momenta in the Brillouin zone
into $N$ divisions. We obtain the BCDW phase near $U=2V$ by
employing this consistent and well-controlled RG method at one
loop level, with no additional interaction terms or {\it ad hoc}
approximations.
The width of this phase increases with $U$, and continues to expand until the
RG procedure breaks down.
We have verified that as $N$ is increased the phase
boundaries become independent of the number of divisions. Our results
confirm that a BCDW phase emerges spontaneously in the EHM at
half-filling and clarify why this result has eluded
earlier standard g-ology and bosonization techniques.
\acknowledgments We thank A. H. Castro Neto and A. Sandvik for
useful discussions and Boston University for financial support.

\end{document}